\documentclass{bmcart}
\usepackage[utf8]{inputenc}
\usepackage{amsmath}
\usepackage{amsthm}
\usepackage{svg}
\usepackage{amsfonts}
\usepackage[noend]{algorithm2e}

\usepackage{pdfpages}

\usepackage{url}
\usepackage{float}
\restylefloat{table}

\usepackage{graphicx} 

\newcommand{\newml}[1]{{#1}}

\renewcommand{\vec}[1]{\boldsymbol{#1}}

\renewcommand{\u}{\vec{u}}
\renewcommand{\v}{\vec{v}}

\newcommand{\w}{\vec{w}}

\newcommand{\ev}[1]{\langle {#1} \rangle}

\newcommand{\distf}[3]{d_{#1}({#2},{#3})}
\newcommand{\dist}[2]{d({#1},{#2})}

\newcommand{\cnp}{\textsf{CNP-transformation}}

\newtheoremstyle{compactthm}
  {2mm} 
  {2mm} 
  {} 
  {} 
  {\bfseries} 
  {.} 
  {.5em} 
  {} 
\theoremstyle{compactthm}

\newtheorem{theorem}{Theorem}
\newtheorem{lemma}{Lemma}

\newtheorem{proposition}{Proposition}

\begin{document}

\begin{frontmatter}

\begin{fmbox}
\dochead{Research}



\title{Comparing copy-number profiles under multi-copy amplifications and deletions}


\author[addressref={aff1},email={garance.cordonnier-martin-de-gibergues@polytechnique.edu}]{\fnm{Garance} \snm{Cordonnier}}
\author[addressref={aff2},email={manuel.lafond@USherbrooke.ca},corref={aff2}]{\fnm{Manuel} \snm{Lafond}}

\address[id=aff1]{
  \orgname{Department of Computer Science, École polytechnique}, 
  \city{Paris},                              
  \cny{France}                                    
}
\address[id=aff2]{%
  \orgname{Department of Computer Science, Université de Sherbrooke},
  \city{Sherbrooke},
  \cny{Canada}
}

\end{fmbox}


\begin{abstractbox}

\begin{abstract} 

\parttitle{Background}  during cancer progression, malignant cells accumulate somatic mutations that can lead to genetic aberrations.
In particular, evolutionary events akin to segmental duplications or deletions can alter the copy-number profile (CNP) of a set of genes in a genome.  
Our aim is to compute the evolutionary distance between two cells for which only CNPs are known.  This asks for the minimum number of segmental amplifications and deletions to turn one CNP into another.  This was recently formalized into a model where each event is assumed to alter a copy-number by $1$ or $-1$, even though these events can affect large portions of a chromosome. 

\parttitle{Results}  we propose a general cost framework where an event can modify the copy-number of a gene by larger amounts. 
We show that any cost scheme that allows segmental deletions of arbitrary length makes computing the distance strongly NP-hard.  We then devise a factor $2$ approximation algorithm for the problem when copy-numbers are non-zero and provide an implementation called \textsf{cnp2cnp}.  We evaluate our approach experimentally by reconstructing simulated cancer phylogenies from the pairwise distances inferred by \textsf{cnp2cnp} and compare it against two other alternatives, namely the \textsf{MEDICC} distance and the Euclidean distance.

\parttitle{Conclusions} the experimental results show that our distance yields more accurate phylogenies on average than these alternatives if the given CNPs are error-free, but that the \textsf{MEDICC} distance is slightly more robust against error in the data.
In all cases, our experiments show that either our approach or the \textsf{MEDICC} approach should preferred over the Euclidean distance.


\end{abstract}



\end{abstractbox}
%

\end{frontmatter}


\section*{Background}

Cancer is widely recognized as an evolutionary process during which cells within a population accumulate aberrant somatic mutations and replicate indefinitely~\cite{nowell1976clonal}.  
These cells are divided in subpopulations, called \emph{clones}, that share common mutation traits and form tumors.  A natural problem that arises is to reconstruct the evolution of a set of clones within a tumor.  This question has recently led to the development of several phylogenetic algorithms tailored for cancer evolution. 
Most of them use either information of single nucleotide variants obtained from bulk~\cite{jiao2014inferring,el2015reconstruction,malikic2015clonality,yuan2015bitphylogeny} or single-cell~\cite{jahn2016tree,ross2016onconem,el2018sphyr} sequencing data, or copy-number alterations~\cite{navin2011tumour,abo2014breakmer,el2016copy,xia2018phylogenetic,zhou2015maximum} (usually in the context of single-cell data).
We refer the reader to~\cite{schwartz2017evolution} for a survey of these methods.

In this work, we are interested in the problem of inferring the minimum number of copy-number alteration events that explain how a cell evolved into another.  
In tumors, several events can make the copy-number of a gene different from the normal diploid two-copy state, thereby creating \emph{copy-number aberrations}.  
As an example, the \emph{breakage–fusion–bridge} (BFB) phenomenon~\cite{lo2002dna} occurs when a region including a telomere breaks off a chromosome.  During replication, two sister chromatids have unterminated ends and they fuse, leading to what is essentially a chromosome portion concatenated with a reversed copy of itself (see~\cite{lo2002dna,marotta2012common}  for a more thorough explanation).  \newml{Afterwards, the centromeres of the fused chromatids get pulled in opposite directions, leading to another breakage.  This BFB cycle repeats until the chromatids receive a telomere (usually after translocation)}.  Each BFB event potentially doubles the copy-number of a gene, \newml{and since these events are known to occur in cycles, a gene copy-number may become significantly higher than normal (i.e. more than double) in a short evolutionary time span}.
Other examples of events include focal deletions~\cite{rajaram2013two,liu2016deletions} or missegregation of chromosomes~\cite{holland2009boveri}.

Desper et al.~\cite{desper1999inferring} were among the first to consider copy-number aberrations for phylogenetic reconstructions, using comparative genomic hybridization (CGH) data to reconstruct a mutation hierarchy.
In~\cite{liu2009inferring}, Liu et al. propose a distance-based approach based on CGH data to infer multi-cancer phylogenies.  
Single-cell phylogenetics then gained widespread attention in an influential paper of Navin et al.~\cite{navin2011tumour}.  The authors applied single-nucleus sequencing on a breast cancer tumor, obtained the \emph{copy-number profile} (CNP) 
of several cells, each represented as a vector of integers, and used the Euclidean distance to compare two CNPs.  
Later, Schwarz et al.~\cite{schwarz2014phylogenetic} pointed out that 
a single event can amplify or delete large portions of a chromosome, thereby altering the copy-number of several genes and making the Euclidean distance  overestimate the true number of events.  

The authors proposed the following methodology to compare two CNPs.  
\newml{First, assuming diploid genomes, the copy-number for the two alleles of each gene (which can differ) is inferred from sequencing data.  The correspondence between the copy-numbers and the alleles is unknown, so a \emph{phasing} step must be applied.  This consists of assigning each copy-number to one of the two alleles (this is done under a minimum-evolution principle, see~\cite{schwarz2014phylogenetic} for details).
After this step, each chromosome can be represented as a pair of CNPs, and chromosomes from two cells can be compared by computing the distances between the corresponding alleles.  The distance proposed is the minimum number of \emph{segmental} amplification and deletion events required to transform a given CNP into another.}

\begin{figure}
    \centering
    \includegraphics[width=0.6\textwidth]{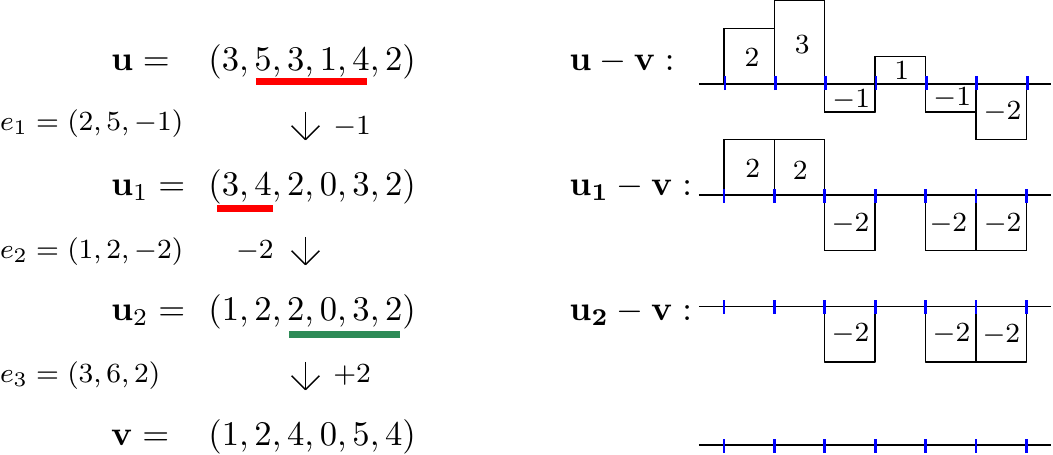}
    \caption{Left: two CNPs $\u$ and $\v$, represented as integer vectors.  The CNP $\u$ can be turned into $\v$ with three events: two deletions and one amplification.  Right: a visual representation of the difference vectors obtained at each step.  Note that a $0$ remains a $0$ even after amplification.}
    \label{fig:fig1}
\end{figure}

\newml{In this work, we focus on the latter step.
We assume that the CNP inference and the phasing steps have been performed, and must find a most parsimonious sequence of events explaining two given CNPs}.  This is analogous to  classical rearrangements problems~\cite{fertin2009combinatorics}, but the main novelty (and difficulty) of CNP comparison is that only copy-numbers are known, not the ordering of genes.  In~\cite{schwarz2014phylogenetic}, Schwarz et al. introduced the \textsf{MEDICC} model, which approximates  segmental events on a chromosome by events that alter a subinterval of a CNP by +1 or -1.  
Figure~\ref{fig:fig1} shows an example turning a CNP $\u$ into another $\v$ (under our model where any amount of change is allowed).
The problem of computing the minimum number of subinterval alterations to transform one CNP into another
was solved in exponential-time in~\cite{schwarz2014phylogenetic} by modeling CNP events with a finite-state transducer. Zeira et al.~\cite{zeira2017linear} gave a linear time algorithm, using a clever trick for computing each row of a quadratic-size dynamic programming table in constant time (similar to the techniques used in~\cite{lafond2012optimal}).
In~\cite{el2016copy}, the large phylogeny problem under this model is shown NP-hard, though solvable with an ILP.  They also present the \emph{copy-number triplet} problem, which when given two CNPs $\u$ and $\v$ asks for a CNP whose sum of distances to $\u$ and $\v$ is minimized.  The problem can be solved in pseudo-polynomial time $O(n^2N^7)$, where $n$ is the CNP size and $N$ the maximum copy number. 
Other distances and phylogenetic approaches are discussed in~\cite{xia2018phylogenetic,abo2014breakmer,zhou2015maximum,el2016copy,letouze2010analysis,qingge2018minimum}

\noindent
\textbf{Our results.}  The above CNP comparison frameworks limit events to alter copy-numbers by $1$ or $-1$.  As we exemplified with BFB, several copies of a gene can be affected by a single event.  Moreover, the MEDICC software has a copy-number limit of $4$, making it inappropriate for genes attaining copy-numbers in the tens, twenties or even more, as has been reported for e.g. the MYC or EGFR genes~\cite{santarius2010census,park2014high,campbell2017association}.
In this work, we address these limitations by generalizing the Copy-Number Transformation problem defined in ~\cite{schwarz2014phylogenetic,zeira2017linear}.  
We define a distance $\distf{f}{\u}{\v}$ between two CNPs $\u$ and $\v$ 
which assigns a weight of $f(c, \delta)$ to an event that alters a copy-number of $c$ by an amount of $\delta$.
We show that computing $\distf{f}{\u}{\v}$ becomes strongly NP-hard whenever we allow deletions of any amount at unit cost.  
In the context of our problem, ``strongly'' means that our hardness holds even if $N$, the maximum value in $\u$ and $\v$, is polynomial in $n$, the number of elements in our CNPs.  This is especially relevant, given that the \textsf{MEDICC} model was initially solved in time $O(nN)$ and that the copy-number triplet problem can be solved in time $O(n^2N^7)$.  Our result implies that such pseudo-polynomial time algorithm are impossible in our case unless P = NP. 
We then show that if any amount of change is permitted across an interval at unit cost, 
then a simple linear-time factor $2$ approximation algorithm can be devised.
We validate our approach by reconstructing simulated phylogenies using neighbor-joining (NJ), and compare them with the \textsf{MEDICC} distance and Euclidean distance.  \newml{We perform our experiments on error-free data and noisy data (where the true copy-numbers are altered by a random amount).  Using a variety of simulation papameters, we show that both our distance and the \textsf{MEDICC} distance achieve significantly better accuracy than the Euclidean distance.  Our distance is slightly more accurate on error-free data, and the \textsf{MEDICC} distance is slightly more tolerant to error.}

\section*{Results}

We first provide the required preliminary notions required to state our theoretical results.  We then show that computing our copy-number distance is strongly NP-hard, and present our approximation algorithm.  Finally, we present our experimental results on reconstructing simulated phylogenies.

\subsection*{Preliminary notions}

Throughout the paper, we use the interval notations $[n] = \{1, 2, \ldots, n\}$ and $[s,t] = \{s, s+1, \ldots, t\}$.  Given a vector $\u = (u_1, \ldots, u_n)$ of $n$ integers and $i \in [n]$, we will always write $u_i$ for the value at the $i$-th position of $\u$.  If $u_i = 0$, then $i$ is called a \emph{null} position.
We will assume that every vector $\u$ of dimension $n$ has special values $u_0 = u_{n+1} = 0$.
We denote by 
$\u^{-\{i\}}$ the vector obtained by removing position $i \in [n]$, 
i.e. $\u^{-\{i\}} = (u_1, \ldots, u_{i-1}, u_{i+1}, \ldots, u_n)$.
If $\v$ is a vector of the same dimension, then $\u - \v = (u_1 - v_1, \ldots, u_n - v_n)$.

We assume that a reference chromosome is partitioned into contiguous subsequences, called \emph{positions}, each numbered from $1$ to $n$.  A \emph{copy-number profile} (CNP)  is a vector $\u = (u_1, \ldots, u_n)$ of non-negative integers representing the copy-number of each position in a clone.  
We consider \emph{amplification} and \emph{deletion} events, which respectively have the effect of increasing and decreasing the number of copies in a chromosome.  As in~\cite{schwarz2014phylogenetic,zeira2017linear}, we assume that events affect a set of positions that are contiguous in the reference chromosome.

An \emph{event} is a triple $e = (s, t, \delta)$ where $1 \leq s \leq t \leq n$ and $\delta \in \mathbb{Z} \setminus \{0\}$.  Here the $[s, t]$ interval depicts the set of affected positions, and $\delta$ is the amount of change.  The event $e$ is an amplification when $\delta > 0$ and a deletion when $\delta < 0$.  A copy-number cannot drop below $0$ and cannot increase from a $0$ to another value (e.g. new genes cannot be created once completely lost).
Applying event $e = (s,t,\delta)$ on a CNP $\u$ yields another CNP $\u' = (u'_1, \ldots, u'_n)$ with, for $i \in [n]$, 

\vspace{-7mm}

\[
u'_i = 
\begin{cases}
\max(u_i + \delta, 0) &\mbox{if $i \in [s, t]$ and $u_i > 0$} \\
u_i &\mbox{otherwise}
\end{cases}
\]

\vspace{-2mm}

We denote by $\u\ev{e}$ the CNP obtained by applying event $e$ on a CNP $\u$.  More generally, if $E = (e_1, \ldots, e_k)$ is an ordered sequence of events, we write $\u\ev{E} = \u\ev{e_1}\ev{e_2}\ldots\ev{e_k}$ to denote the CNP obtained by applying each event of $E$ in order.  We may also write $\u\ev{e_1 \ldots e_k}$ instead of $\u\ev{(e_1, \ldots, e_k)}$.  Given two CNPs $\u$ and $\v$ of dimension $n$, we say that $E$ \emph{transforms} $\u$ into $\v$ if $\u\ev{E} = \v$.

We will often use the \emph{difference vector} of $\u$ and $\v$, and usually denote $\w := \u - \v$.  The representation of $\w$ as in Figure~\ref{fig:fig1} on the right provides the following intuition: if $\u \ev{E} = \v$, then the events of $E$ need to ``squish'' that values of $\w$ to $0$ to make $\u$ equal to $\v$ (ensuring that no value $u_i$ of $\u$ drops to $0$ in the process unless $v_i = 0$).

\subsubsection*{Minimum cost transformations}


\noindent
Given two CNPs $\u$ and $\v$, our goal is to find a minimum-cost sequence $E$ that transforms $\u$ into $\v$.  
In~\cite{schwarz2014phylogenetic,zeira2017linear}, the cost of an event $(s,t,\delta)$ is $|\delta|$. 
Here, we propose a generalization by defining a cost function $f: \mathbb{N} \times \mathbb{Z} \rightarrow \mathbb{N}^{>0}$ that assigns a positive cost to altering a copy-number $c$ by an amount of $\delta$.
That is, if we apply $(s, t, \delta)$ on $\u$, each position $i \in [s,t]$ has its own corresponding cost $f(u_i, \delta)$, which could be interpreted as the plausibility of going from copy-number $u_i$ to $max(u_i + \delta, 0)$.  We then define the cost $cost_f(\u, e)$ with respect to $f$ of applying $e = (s,t,\delta)$ on $\u$ as the maximum cost within $[s,t]$, i.e.

$$cost_f(\u, e) = \max_{i \in [s,t]} f(u_i, \delta)$$

The events proposed in the MEDICC algorithm of Schwarz et al.  can be decomposed into $\delta$ events of unit cost.  This can be modeled under our framework with a function $mdc$ defined as
$mdc(u_i, \delta) = 1$ if $\delta \in \{-1, 1\}$ and 
$mdc(u_i, \delta) = \infty$ otherwise.
%
%
%
%
Alternatively, one could state that a position with copy-number $u_i$ can hardly more than double in a single event (assuming that amplifications are duplications), but that deletions can suppress any number of copies.  We call this the doubling function $dbl$, defined as $dbl(u_i, \delta) = 1$ if $u_i + \delta \leq 2u_i$, and $dbl(u_i, \delta) = \infty$ otherwise.



Finally, the most permissive cost function $any$ allows any movement without  constraint: 
simply define $cost(u_i, \delta) = 1$ for any $\delta \in \mathbb{Z}$.  \newml{This can, for instance, be used to model succession of events that can potentially amplify copy-numbers above their double in a short time span --- an example of this being BFB cycles.}


In this paper, we mostly analyze the $any$ function for its simplicity, but will sometimes use the $dbl$ function for its relevance.
Given two CNPs $\u$ and $\v$ and a cost function $f$, the \emph{cost} of a sequence of events $E = (e_1, \ldots, e_k)$ satisfying $\u\ev{E} = \v$
is equal to the sum of the cost of applying successive events of $E$ on $\u$, i.e. 

\vspace{-8mm}

\[
cost_f(\u, E) = cost_f(\u, e_1) + cost_f(\u\ev{e_1}, e_2) + \ldots + cost_f(\u\ev{e_1, \ldots, e_{k-1}}, e_k)
\]

\vspace{-2mm}

If $cost_f(\u, E) \leq cost_f(\u, E')$ for any other sequence $E'$ satisfying $\u\ev{E'} = \v$, then $E$ is called \emph{optimal}.
The $f$-distance between $\u$ and $\v$, denoted $\distf{f}{\u}{\v}$, is the cost of an optimal sequence of events transforming $\u$ into $\v$.
Observe that this ``distance'' is not symmetric (hence the use of double-quotes).  For instance, if $\u = (1,1)$ and $\v = (0, 0)$, then $\distf{mdc}{\u}{\v} = 1$ but $\distf{mdc}{\v}{\u}$ is undefined since $\v$ cannot be transformed into $\u$.  
We will therefore usually assume that $\u$ does not have any null position.
We note here that the \emph{median} distance, defined as $\min_{\w \in V} (\distf{f}{\w}{\u} + \distf{f}{\w}{\v})$ (where $V$ ranges over $\mathbb{Z}^n$), is symmetric for all the functions mentioned above.  However, no efficient algorithm is known for any median distance. 
Our problem is the following.

\vspace{2mm}

\noindent
The \cnp~problem:

\noindent
\textbf{Given:} a source CNP $\u$, a target CNP $\v$, a cost function $f$ and an integer $k$;

\noindent
\textbf{Question:} is $dist_f(\u, \v) \leq k$?

\vspace{2mm}

We say that $f$ is a \emph{unit-cost} function if $f(c, \delta) \in \{1, \infty\}$ for any $c$ and $\delta$ (e.g. the functions $mdc, dbl$ and $any$).  
A cost function $f$ is called \emph{deletion-permissive} if $cost_f(u_i, \delta) = 1$ for any $u_i$ and any $\delta < 0$, i.e. there is no particular constraint on deletions.  We will mainly focus deletion-permissive functions, the rationale being that unlike duplications, deletions could suppress an arbitrary number of copies.  

\subsubsection*{General properties}


\noindent
Before proceeding with our results on computing $f$-distances, we present some results of general interest that will be useful later on.

\begin{proposition}\label{prop:remove-stuff}
For any two CNPs $\u$ and $\v$ of the same dimension, any position $i \in [n]$ and any cost function $f$, $\distf{f}{\u}{\v} \geq \distf{f}{\u^{-\{i\}}}{\v^{-\{i\}}}$
\end{proposition}

We omit the proof details.  The idea is that given a sequence of events $E$ transforming $\u$ into $\v$, we can apply $E$ on $\u^{-\{i\}}$ by ignoring position $i$ when it is affected.
A sequence of events $E$ is called \emph{amp-first} if all amplifications appear before all deletions.  An \emph{amp-first reordering} of a sequence $E$ is an amp-first sequence $E'$ 
that contains the same events as $E$.  Notice that if $E$ has $a$ amplifications and $d$ deletions, then there are $a!d!$ amp-first reorderings of $E$.

\begin{figure}[b]
    \centering
    \includegraphics[width=0.95\textwidth]{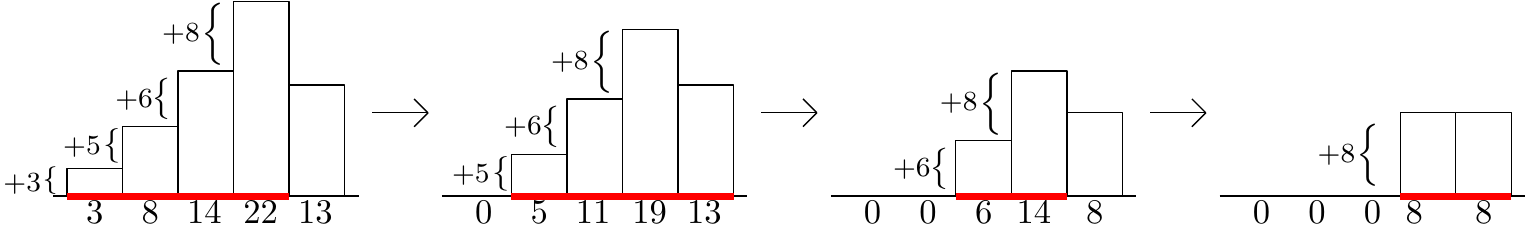}
    \caption{A visual representation of the difference vector $\w = \u - \v$ leading to a staircase of length $4$ in interval $[1,4]$.  For instance, setting $\v = (1,1,1,1,1)$ and $\u = (4,8,15,23,14)$ would lead to the situation shown above.  A smooth deletion sequence turning $\u$ into $\v$ is shown (last deletion omitted).
    }
    \label{fig:staircase}
\end{figure}

\begin{proposition}\label{prop:any-order}
Let $\u$ and $\v$ be two CNPs with no null positions.  If a sequence $E$ satisfies $\u\ev{E} = \v$, then any amp-first reordering $E'$ of $E$ satisfies $\u\ev{E'} = \v$.
\end{proposition}

\vspace{-7mm}

\begin{proof}
Denote $E = ((s_1, t_1, \delta_1), \ldots, (s_k,t_k,\delta_k))$.  
For any position $i$, the sum $\sum_{j=1}^k \delta_k$ does not change even if we reorder the events in $E$, so $u_i$ should still become $v_i$ after reordering the event and applying them on $\u$.
The only danger is that a position drops to $0$ since $\v$ has no null position, but this cannot happen if all amplifications are moved in front of $E$.
\end{proof}

\vspace{-5mm}

Given a CNP $\w$ of length $n$, an interval $[a, b]$ is a \emph{staircase of $\w$} if $0 < w_a < w_{a+1} < \ldots < w_b$.
The \emph{length} of the staircase $[a, b]$ is $b - a + 1$.  Figure~\ref{fig:staircase} depicts a staircase of length $4$.
The next lemma can be useful to obtain quick lower bounds on a particular instance, and plays an important role in our hardness result (proof in Supplementary material).

\begin{lemma}\label{lem:staircase}
Let $\u, \v$ be two CNPs with no null positions.
If $\u - \v$ contains a staircase $[a, b]$ of length $k$, then $\distf{f}{\u}{\v} \geq k$ for any unit-cost function $f$.
\end{lemma}

\vspace{-4mm}

\subsection*{Strong NP-hardness}

We show that the \cnp~problem is \emph{strongly} NP-hard.  
This result holds for any deletion-permissive unit-cost function $f$, and even if $\u$ and $\v$ contain no null position (we note that in~\cite{zeira2017linear}, null positions make the problem more complex, but not here).  In particular, the hardness also holds if only deletions are allowed.
We assume that we are given two CNPs $\u$ and $\v$ and we put $\w := \u - \v$.  

Suppose that $\w$ contains a staircase in interval $[1, k]$ for some $k$, and that $\distf{f}{\u}{\v} = k$. A sequence $E = (e_1, \ldots, e_k)$ such that $\u\ev{E} = \v$ is called \emph{smooth} if, for every $i \in [k]$, $e_i = (i, b_i, w_{i-1} - w_{i})$ for some $b_i \geq k$.  Intuitively, $E$ removes the first step, then the second, and so on, see Figure~\ref{fig:staircase}.
Observe that in a smooth deletion sequence, the positions to the right of $k$ may or may not be affected by deletions.

\begin{lemma}\label{lem:smooth-stairs}
Let $\u$ and $\v$ be two CNPs with no null positions and let $f$ be any unit-cost function.  If $\u - \v$ contains a staircase in interval $[1, k]$ and $\distf{f}{\u}{\v} = k$, then there exists a smooth sequence transforming $\u$ into $\v$.
\end{lemma}

Lemma~\ref{lem:smooth-stairs} requires the most technical proof of the paper (by far), and we defer it to the Supplementary material.
The reduction becomes relatively simple when given this lemma.
%
%
Our reduction is from the  \textsf{3-partition} problem.
In this problem, we are given a multi-set $S = \{s_1, \ldots, s_n\}$ of $n = 3m$ positive integers.  Defining $t := \frac{1}{m} \sum_{i \in [n]}s_i$, we are asked whether $S$ can be partitioned into $m$ subsets $S_1, \ldots, S_m$, each of size $3$, such that $\sum_{s \in S_i} s = t$ for all $i \in [m]$.
This problem is known to be strongly NP-hard~\cite{garey2002computers} (i.e. it is hard even if the values of $S$ are $O(n^k)$ for some constant $k$).
The proof can be found in the Supplementary material.

\begin{theorem}\label{thm:hardness}
The \cnp~problem is strongly NP-hard for any deletion-permissive unit-cost function, even if the CNPs have no null positions.
\end{theorem}


\subsection*{Approximation algorithm}

In this section, we show that if $\v$ does not contain any null position, then $\distf{any}{\u}{\v}$ can be approximated within a factor of $2$ in linear time.  We discuss practical ways of handling null positions at the end of the section.  We now assume that $f = any$ and will write $\dist{\u}{\v}$ instead of $\distf{any}{\u}{\v}$.

As usual, $\u$ and $\v$ are the source and target CNPs, respectively, and $\w := \u - \v$.  
The idea of the approximation is that 
if two consecutive positions $i$ and $i + 1$ have the same difference between $\u$ and $\v$, i.e. $w_i = w_{i+1}$, then their value needs to change by the same amount.  It might then be a good idea to treat these positions as one and always affect both with the same events.  In fact, a whole interval of equal $\w$ values can be treated as a single position.  We show that the number of distinct equal intervals gives a good bound on $\dist{\u}{\v}$.

\subsubsection*{Approximation by flat intervals}


\noindent
Recall that if $\w$ is a vector of $n$ integers, it has implicit values $w_0 = w_{n+1} = 0$.  We say that $[a,b]$, with $0 \leq a \leq b < n + 1$, is a \emph{flat interval} if $w_i = w_j$ for every $a \leq i, j \leq b$.  
If no interval properly containing $[a,b]$ is flat, then $[a,b]$ is a \emph{maximal} flat interval. 
In fact, in the remainder, we will omit the term ``maximal'' and always assume that discussed flat intervals are maximal.
We write $F_{\w}$ for the set of flat intervals of $\w$.  
Note that this set is well-defined and that it partitions $[0, n+1]$, by the maximality property.  
The intervals that contain $0$ and $n+1$ in $F_{\w}$ are called \emph{extreme flat intervals}, and always have a value of $0$ (also, these intervals are possibly $[0,0]$ and/or $[n+1, n+1]$, but not necessarily).
The key lemma says that $\distf{f}{\u}{\v}$ is at least about half the number of flat intervals (see Supplementary material).

\begin{lemma}\label{lem:fovr2}
Let $\u, \v$ be two distinct CNPs with no null positions, and let $\w := \u - \v$. 
Then for any unit-cost function $f$, 
$\distf{f}{\u}{\v} \geq \lceil (|F_{\w}| - 1)/2  \rceil$.

\end{lemma}

Lemma~\ref{lem:fovr2} yields a very simple factor 2 approximation algorithm: compute $F_{\w}$, and return $|F_{\w}| - 2$.  This corresponds to a solution in which we treat each flat interval separately (ignoring the two extremities) and is guaranteed to be at most twice the optimal number of events.
Computing $F_{\w}$ can be done in a single pass through $\w$ by increasing a counter whenever we encounter a position $i$ with $w_i \neq w_{i-1}$.

\begin{theorem}
The \cnp~problem can be approximated within factor $2$ in linear time for cost function $f = any$ when the CNPs contain no null position.
\end{theorem}

It is open whether this could be adapted to other functions, e.g. the $dbl$ function.

\subsubsection*{Improvements to the approximation algorithm}


\noindent
We first observe that the bound in Lemma~\ref{lem:fovr2} is essentially tight.  This can be seen with any $\u, \v$ such that 
$\u - \v = (1,2,3, \ldots, k-1, k, k-1, \ldots, 3,2,1)$ for some $k$.  Indeed, one can decrease $|F_{\w}|$ by two at each round.
On the other hand, our naive $2$-approximation is twice as bad as optimal. 
We show how to improve this in a heuristic fashion by devising an algorithm that can only perform better than the naive one.  We leave it as an open problem to determine the approximation guarantees of this  algorithm.

Our goal is to apply events that reduce $|F_{\w}|$ by two as many times as possible.
In a greedy fashion, we apply the following strategy for our improved $2$-approximation:
as long as $\u \neq \v$, find an event $e$ that reduces $|F_{\w}|$ by $2$, if one exists, and apply it to $\u$. If no such event exists, take the leftmost non-extreme flat interval $[a,b]$ of $\w$ and apply the event $(a,b,-w_a)$.  Repeat until $\u = \v$.  

An event $(i,j, \delta)$ reduces $|F_{\w}|$ by $2$ precisely when $w_{i-1} - w_i = w_{j+1} - w_j = \delta$ ($i = j$ is possible).  This way we can merge the two flat intervals at the ends of $[i,j]$.  
One can find a good interval by checking all the $O(n^2)$ subintervals $[i,j]$ and then, for each of them, checking whether $w_{i-1} - w_i = w_{j+1} - w_j$.  Moreover, we must check whether applying the event $(i,j,\delta)$ would make a value of $\u$ go below $0$.  Verifying every possible event can be done in time $O(n^3)$ and as there are $O(n)$ flat intervals, the algorithm takes time $O(n^4)$.


This can be improved to $O(n^2 \log n)$ by finding good events in time $O(n \log n)$.  Due to space constraints, we relegate the detailed analysis of the improved heuristic to the Supplementary material.  The idea is to scan $\w$ from left to right and store in a \emph{treap} data structure (see~\cite{seidel1996randomized}) the set of flat intervals encountered so far, which allows to detect quickly whether the current flat interval could be matched with another one.

\subsubsection*{Handling null positions}


\noindent
Our approximation ratio is not guaranteed to hold when there are many null positions.
However, we show that in many practical cases, we can simply ignore null positions and remove them.
In particular, we may assume that $\v$ has no two consecutive null positions
(Lemma~\ref{lem:conseq-zero}) and that for any null position $i$ in $\v$, 
we have $w_{i-1} < w_i$ and $w_{i+1} < w_i$ (Lemma~\ref{lem:outstanding-nulls}).
Thus instances with null positions can be reduced to ones where 
the only null positions remaining are ``sandwiched'' between non-null positions with a smaller value in $\w$.

Note that our approximation can still perform badly with these two conditions.  For instance, suppose that $\u = (15,2,15, 2, \ldots,15,2)$ and $\v = (14,0,14,0, \ldots, 14, 0)$.  
We would solve this in about $n/2$ events.  However, the two events $(1, n, -2), (1, n, 1)$ turn $\u$ into $\v$.  Designing a better approximation for these cases is an open problem.

\begin{lemma}\label{lem:conseq-zero}
Suppose that $v_i = v_{i+1} = 0$ for some position $i$.  Then removing position $i$ or $i + 1$, whichever is smaller in $\u$, from $\u$ and $\v$ preserves the distance between $\u$ and $\v$.  
Formally, for any unit-cost function $f$, if $u_i \geq u_{i+1}$, then $\distf{f}{\u}{\v} = \distf{f}{\u^{-\{i + 1\}}}{\v^{-\{i + 1\}}}$.  Similarly if $u_{i+1} \geq u_i$, then 
$\distf{f}{\u}{\v} = \distf{f}{\u^{-\{i\}}}{\v^{-\{i\}}}$.
\end{lemma}

\begin{lemma}\label{lem:outstanding-nulls}
Suppose $v_i = 0$ for some position $i$ and that
$w_{i - 1} \geq w_i$ or $w_{i + 1} \geq w_i$.  Then $\distf{f}{\u}{\v} = \distf{f}{\u^{-\{i\}}}{\v^{-\{i\}}}$ for any unit-cost function  $f$.
\end{lemma}

\subsection*{Experiments}

We tested our flattening approximation algorithm and its improved version on simulated chromosomes that evolve along a tree through segmental tandem duplications and losses.  Chromosomes were represented as strings of genes.
Note that we did not simulate CNP evolution under the assumptions of our model.  We evolved actual sequences as opposed to integer vectors, and the initial ordering of genes could be broken after several events.  
Our goal was to reconstruct phylogenies from the distances between the CNPs of the chromosomes at the leaves of the tree. We used the NJ implementation in Phylip~\cite{saitou1987neighbor, felsenstein1993phylip} and compared four distances: (1) our improved approximation; (2) our flat interval count; (3) the $mdc$ cost, as in the \textsf{MEDICC} model; and (4) the Euclidean distance.
To compute $d_{mdc}$, we implemented the dynamic programming algorithm of Zeira, Zehavi and Shamir~\cite{zeira2017linear}, hereafter called the ZZS algorithm (we could not use the MEDICC software as it only handles copy-numbers up to $4$).  The Euclidean distance is defined as $\sqrt{\sum_{i=1}^n (u_i - v_i)^2}$, as used in~\cite{navin2011tumour}.  For the first three distances, we took the minimum of $d(\u, \v)$ or $d(\v, \u)$ to get a symmetric distance, removing null positions of $\u$ and filtering null positions of $\v$ as in Lemmas~\ref{lem:conseq-zero} and~\ref{lem:outstanding-nulls}.


\subsubsection*{Simulated tree generation}

    We now describe how the trees were generated.
    First, we select a rooted binary tree $T$ on $l$ leaves labeled $\{1, \ldots, l\}$ uniformly at random.  This is achieved by using the recursive splitting process described by Aldous in~\cite{aldous1996probability}, which starts with a completely unresolved tree, splits the root in two subtrees chosen uniformly at random, and repeats on these subtrees.
    We then assign to the root $r$ of $T$ an \emph{exemplar chromosome}, i.e. any string in which each gene occurs exactly once (note that the initial ordering of genes does not matter for our purposes).
    
    Then for each branch $uv$ of $T$ from top to bottom, we select a random number of events $k$ chosen uniformly at random in the interval $[e_{min}, e_{max}]$, where $e_{min}, e_{max}$ are simulation parameters.  To introduce some rate heterogeneity among branches, we then multiplied $k$ by a random number chosen from a uniform distribution with mean and  standard deviation $1$.
    The chromosome string at node $v$ is obtained by applying $k$ random events on the chromosome string associated with its parent $u$.  
    Each event is either a tandem duplication with probability $\Delta$ or a deletion with probability $1 - \Delta$.  The starting position of each event is chosen uniformly at random on the chromosome string and, to find the length $t$ of the substring affected, we apply the following process. Start with $t = 1$, then apply the following: as long as a random number between $0$ and $1$ is above a given parameter $r$, increment $t$ by $1$ and repeat.  We stop at the first random number below $r$.  We chose to consider only values $r \leq 0.1$ since higher values resulted in copy-numbers in the hundreds, sometimes even in the thousands, a road which we did not deem necessary to explore.  Setting $r$ between $0.01$ and $0.1$ generally resulted in copy-numbers inside $[0, 50]$. We also note that we also experimented on a model where the event length was chosen as a random fraction of the chromosome length --- this led to exponential copy-number growth and we did not investigate this model further.

    We observed that this process had a tendency to produce leaf chromosomes with CNPs having between 50-60\% null positions under most parameter combinations. 
    This might be deemed unrealistic, and
    furthermore, our results show that no method is able to predict accurate trees under these conditions.
    To avoid this, we added a condition in the loop determining the length $t$ of an event: if incrementing $t$ implies deleting the last occurrence of a gene, we continue the procedure with probability $q$ and stop with probability $1 - q$ (where $q$ is another simulation parameter).  This can be seen as modeling the idea that there may be resistance when attempting to remove every copy of a gene required for survival.  Using $q$ parameter values $0.25, 0.5$ and $0.75$, the proportion of null positions stayed in the intervals 2-5\%, 6-10\% and 15-25\%, respectively.

Note that since each possible tree on $l$ leaves is equally likely to be chosen, the root-to-leaf distances in a tree can be significantly different, and hence the trees are not expected to be ultrametric (for instance, a caterpillar can be selected as well as a perfectly binary tree).  

Since it is difficult to determine the most realistic simulation conditions, we tested several combinations of parameters for the generation of phylogenies.  The summary of the simulation parameters, along with their possible and default values, are summarized here:

\begin{itemize}
    \item 
    $l \in \{10, 50, 100\}$ is the number of leaves in the tree.  The default is $l = 100$;
    
    \item 
    $n \in \{10, 100, 250\}$ is the number of genes (i.e. distinct characters) in the root chromosome ($n$ is also the number of positions in our vectors).  The default is $n = 100$;
    
    \item
    $(e_{min}, e_{max}) \in \{(2,4), (5,10), (20,40)\}$ is the range of the possible number of events on each branch.  The default is $(e_{min}, e_{max}) = (5, 10)$;
    
    \item
    $\Delta \in \{0.25, 0.5, 0.75\}$ is the 
    probability that an event is a duplication (and $1 - \Delta$ the probability that an event is a loss).  The default is $\Delta = 0.5$;
    
    \item 
    $r \in \{0.01, 0.05, 0.1\}$ controls the length of each event: $r$ is the probability that we stop extending the event length.  The default is $r = 0.05$;
    
    \item 
    $q \in \{0.25, 0.5, 0.75, 1\}$ is the probability that a deletion suppresses the last copy of a gene during the length extension procedure (i.e. $1 - q$ is the probability that the extension stops if it would make a copy-number $0$).  The default is $q = 0.25$.
\end{itemize}

\subsubsection*{Tree reconstruction and performance measure}

We generated 50 trees for each parameter combination of interest.  For each tree, we took the chromosome strings at the leaves, obtained their CNPs and provided them as input to each of the four evaluated methods.
We used the normalized Robinson-Foulds (RF)  distance as a measure of the performance of each algorithm~\cite{robinson1981comparison}.  That is, for each inferred tree, we compare it with the ``true'' tree by counting the number of clades that are present in one tree but not the other, divided by $2(l-3)$ (the maximum number of clades that can possibly differ, recalling that $l$ is the number of leaves).  This yields a number between $0$ and $1$: the lower the number, the better we consider the reconstruction.

\subsubsection*{Error tolerance}

It should be noted that the above methodology ignores several sources of errors. 
Inferring exact copy-numbers from single-cell sequencing data is a non-trivial task and is still considered an open problem.  The inferred CNPs are therefore expected to be noisy, especially with genes having a high copy-number.  
Moreover, as discussed in~\cite{schwarz2014phylogenetic}, assigning copy-numbers to their corresponding allele is also a difficult problem.  Here, by only considering single-allele chromosomes, we are supposing that the aforementioned phasing step has been performed correctly, whereas copy-number assignments cannot be assumed to be error-free.  

Both of the above problems have the effect of introducing incorrect copy-numbers into the CNPs.  
To account for this, we gave randomly altered CNPs as input to each method.  More specifically, given an error-rate parameter $\alpha$, for each CNP $\u$ and each position $i$ we changed $u_i$ to a value chosen at random from a normal distribution with mean $u_i$ and standard deviation $\alpha \cdot u_i$ (non-integer values were rounded).  We tested parameter values $\alpha \in \{0, 0.1, 0.25, 0.5, 1\}$.


\begin{figure}[t]

  \includegraphics[width=\linewidth]{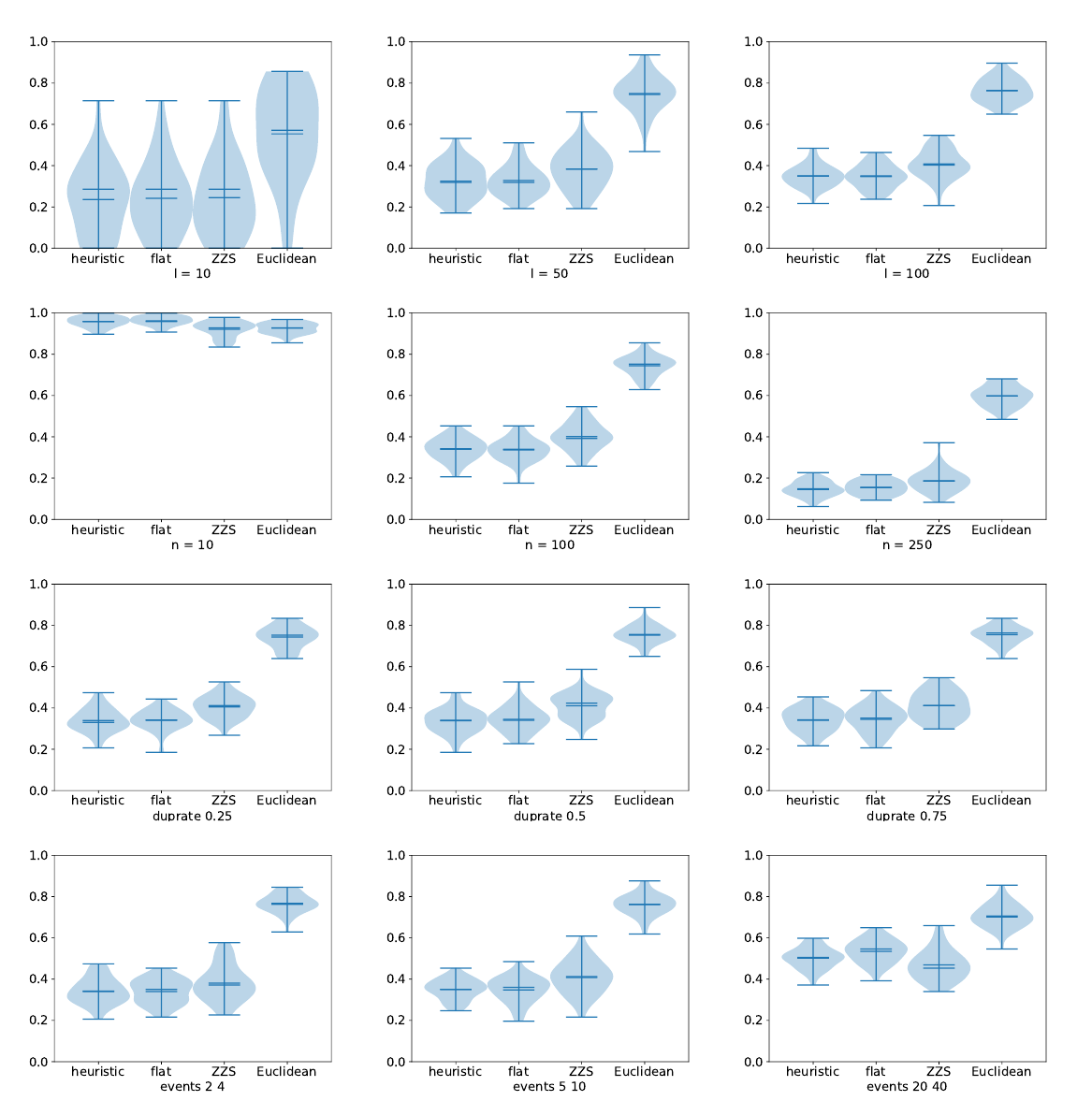}
\caption{Violin plots of the normalized RF distances for our improved approximation (heuristic), the algorithm that counts flat intervals (flat), the ZZS algorithm for the \textsf{MEDICC} model (ZZS), and the Euclidean distance on error-free data.  Each plot summarizes 50 reconstructed trees with, from left to right: (top row) $l = 10, 50$ and $100$ leaves; (second row) $n = 10, 100$ and $250$ genes per chromosome; (third row) duplication rate $\Delta = 0.25, 0.5$ and $0.75$;  (fourth row) possible number of events per branch $(e_{min}, e_{max}) = (2, 4), (5, 10)$ and $(20 ,40)$.
On each row, the other parameters were set to their default as discussed in the text.}
\label{fig:plots}

\end{figure}

\begin{figure}[t]

  \includegraphics[width=\linewidth]{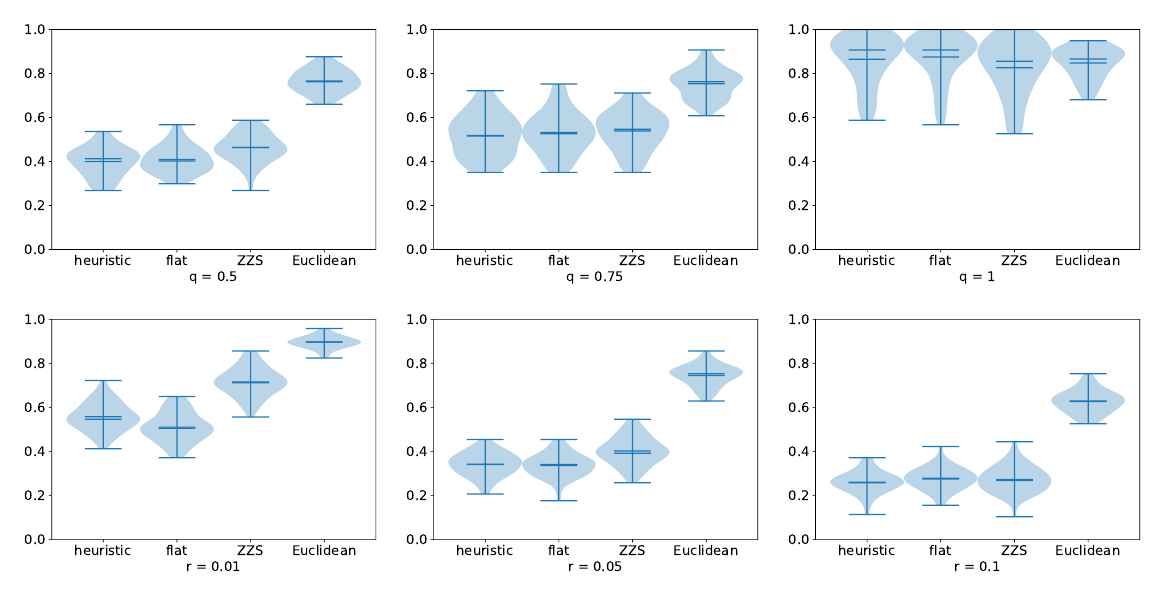}
\caption{Violin plots of the normalized RF distances for the four same approaches with varying parameters $q$ and $r$ (on error-free data).   Other parameters were set to their default values as described in the text.  The plot for $q = 0.25$ is not shown: it is identical to the $n = 100$ plot from Figure~\ref{fig:plots}. }
\label{fig:plots_q}

\end{figure}

\begin{figure}[t]
  \includegraphics[width=\linewidth]{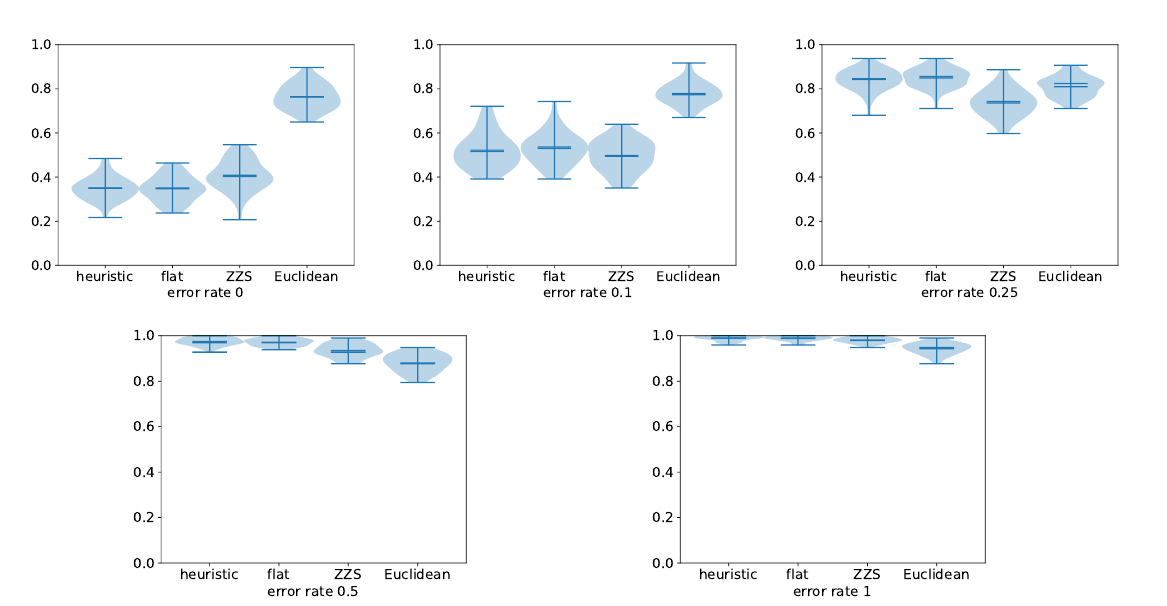}
\caption{Violin plots of the average normalized RF distances for the four same approaches with varying error rates $\alpha \in \{0,0.1,0.25,0.5, 1\}$.  Other parameters were set to their default values as described in the text. }
\label{fig:plots_err}
\end{figure}

\subsubsection*{Experimental results}

We first ran experiments using the default values for all parameters except one in order to isolate the impact of each parameter.
On error-free data, the most interesting results were obtained when varying $l$ and $n$, see Figure~\ref{fig:plots}.  In most situations, our CNP model slightly improves upon the \textsf{MEDICC} model, both of which are significantly better than the Euclidean distance.  The number $n$ of genes is quite important: all the results are poor when each CNP has only $n = 10$ positions, but when $n = 250$, the trees more accurate.  This might be because when $n = 10$, there is not enough opportunity for positions acquire a distinct signature during the evolutionary process, making all distances very similar.  This suggests that many genes or segments should be considered when analyzing copy-number variants in tumor clones.  The duplication rate does not seem to affect the accuracy of the methods, whereas accuracy tends to decrease as the number of events per branch increases.  We do note that when the number of events per branch is within $[5,10]$, our model performs better, but when it is high, i.e. in $[20,40]$, the MEDICC model performs better.  This tendency is confirmed under other parameterizations (see the Supplementary Material).
Figure~\ref{fig:plots_q} show the normalized RF distances when varying parameters $q$ and $r$.  As mentioned before, as $q$ gets closer to $1$, the proportion of null positions is around 50-60\%, making accurate distance computation difficult for all methods. 
As for the $r$ parameter, the accuracy of our approach  is better when $r = 0.01$ but worse when $r = 0.1$.  This tendency can be observed under all parameter combinations (see the Supplementary Material).
One should note that accuracy is generally better if the lengths of events are smaller.
The four approaches on error-free data exhibit similar behavior on other parameter combinations --- additional  plots can be found in the Supplementary Materials.  

The results on CNPs containing errors are summarized in Figure~\ref{fig:plots_err}.  
We observe that the ZZS algorithm achieves slightly more accurate trees whenever the error rate is above zero.  One possible explanation is that a single error in a CNP can split a flat interval into three.  This can significantly alter the flat interval counts, whereas the ZZS distance is less dependant on flat intervals.  The accuracy of the Euclidean distance appears to be the least affected by error rates and even performs better when $\alpha \geq 0.5$.  Observe however that accuracy decays rapidly with error rates: when $\alpha \geq 0.25$, all approaches have an average normalized distance above $0.7$, casting some doubt on their practical usability in this setting.  This suggests that it might be beneficial to apply a CNP error-correction procedure before comparing them (see the Discussion section).
More results on noisy data can be accessed in the Supplementary Material.

To summarize, the heuristic and flat count algorithm always yield a lower average RF distance than the ZZS algorithm on error-free data, except when $n = 10$ (where the average is always above $0.9$ anyways), and every method always outperforms the Euclidean distance. On the other hand, the ZZS approach is slightly more robust to error in the CNP counts.  However, the accuracy of the heuristic, the flat count and ZZS drops quickly as error rates increase.  Even though the Euclidean distance yields better trees at high error rates, their accuracy is still too poor to be able to draw meaningful conclusions from them. 
Whether the \textsf{MEDICC} model is better than ours or not, we believe that either should be preferred over the Euclidean distance when reconstructing phylogenies from distance matrices as in~\cite{navin2011tumour}.

\section*{Discussion}

The results from the experiments section show that 
our CNP distance performs reasonably well on simulated data. 
The incorporation of segmental events into the model does not appear to provide a significant advantage over the unitary events of the ZZS model.  However, the simulations suggest that both approaches yield better results than the traditional Euclidean distance.  This demonstrates that either our method or the ZZS algorithm should be preferred as the CNP comparison component in a single-cell phylogenetic reconstruction pipeline.

It should be noted that our algorithms only approximate the true CNP distance whereas the ZZS algorithm provides an exact solution.  In order to evaluate the true performance of our segmental model, exact approaches should be developed in the future, perhaps using techniques from the field of parameterized complexity.  Moreover, our approaches are very sensitive to errors, even more so than ZZS/MEDICC. 
One possible explanation for this is that both of our algorithms derive their distance from the number of flat intervals.  A single error in a copy-number can turn one flat interval into three, and thus even moderate levels of noise can lead to highly incorrect predictions. 
We believe that the ZZS algorithm is less sensitive to such errors because that, if copy-number differences are large enough, a small error only increases the distance by $1$, which is small in comparison to all the unit events required to handle the high difference.  Therefore, even if the true event distance is overestimated, in a comparative setting the relative distances might be closer to the truth.
It will be interesting to consider CNP error correction procedures based on flat intervals.  
For instance, when performing analysis of multiple cells, one could detect a potentially incorrect copy-number of a given segment by checking whether, after altering a predicted copy number by a small amount, several flat intervals get ``fixed'' when comparing the cell with others.

Another point of interest is that current approaches, including ours and ZZS/MEDICC, ignore rearrangements that change the ordering of segments.  Our models assume that the set of contiguous segments remains the same in all cells during evolution.  However, when duplications and deletions occur, the relative ordering of genes changes and the set of contiguous genes affected by the events will differ from that in the reference.  Inversions, translocations or even chromothripsis also have the same effect. 
This is a difficult problem to handle if only CNPs are known, since integer vectors do not contain enough information to determine which genes are contiguous or not.  One possibility is to ask the following: given two CNPs $C_1$ and $C_2$ to compare, choose a genome $G_1$ whose CNP is $C_1$ and a genome $G_2$ whose CNP is $C_2$ such that the rearrangement distance between $G_1$ and $G_2$ is minimized.

Our work also leaves several questions open.  From a theoretical perspective, it remains to achieve a constant factor approximation when null positions are present in the input.  
Moreover, it is unknown whether the $dbl$ function admits good approximation algorithms and, more generally, whether there are other biologically plausible functions that should be studied.  
On another note, it might be interesting to investigate the copy-number triplet problem (see the introduction) under our model, as it allows to define a symmetric distance between CNPs.  

On a practical level, phylogenetic approaches that are not distance-based should be investigated.   For instance, we could consider maximum parsimony as in~\cite{el2016copy}, where the objective is to minimize the number of events across branches of a tree.  The recent  distance-based approach of Xia et al.~\cite{xia2018phylogenetic}, which is based on the MEDICC model but with an extra error correction step, should also be evaluated in our setting.  
On another note, it remains to test our approach on real data.  
We have ignored the problem of calling copy-numbers and the aforementioned problem of phasing.  These can introduce noise in the data and, as shown in our experiments, all the evaluated methods are sensitive to errors.  This motivates the need for new methods to assign copy-numbers to alleles under our model.
Also, our CNP comparison framework assumes a single-cell setting, where the CNP of each individual cell is known.  Since bulk sequencing is still commonplace, it will be useful to develop methods that are able to compare genomes extracted from samples that contains multiple types of cells.

\section*{Conclusion}

In this work, we provided a general framework for the comparison of CNPs depicting genomes that evolve by segmental amplifications and deletions.  
We have shown that if there is no bound on the number of copies that a deletion can affect, then computing the minimum number of events transforming one CNP into another is strongly NP-hard.  One important implication of this result is that unless P = NP, one cannot use the fact that copy-numbers are not too large (e.g. under $100$) to devise a practical pseudo-polynomial time algorithm, and other solutions must be explored.  On the other hand, we proposed two simple and fast approximation algorithms that were shown to perform reasonably well on simulated datasets.

\subsection*{List of abbreviations}

BFB: breakage-fusion-bridge; CGH: comparative genomic hybridization; CNP: copy-number profile; NJ: neighbor-joining; RF: Robinson-Foulds; ZZS: Zeira, Zehavi, Shamir

\subsection*{Figure listing}

Figure~\ref{fig:fig1}: an example of a CNP-to-CNP transformation.

Figure~\ref{fig:staircase}: a visual representation of a staircase and a smooth deletion sequence.

Figure~\ref{fig:plots}: average normalized RF distances of the four methods evaluated when varying $l, n, \Delta$ and $(e_{min}, e_{max})$.

Figure~\ref{fig:plots_q}: average normalized RF distances of the four methods evaluated when varying $q$ and $r$.

Figure~\ref{fig:plots_err}: average normalized RF distances of the four methods evaluated with varying error rates.

\section*{Declarations}

\begin{backmatter}

\section*{Competing interests}
  The authors declare that they have no competing interests.

\section*{Author's contributions}

GC and ML both participated in writing the manuscript, establishing the theoretical results, performing the experiments and implementing the algorithms.  All authors have read and approved the manuscript.

\section*{Funding}

Publication was funded by the Natural Sciences and Engineering Research Council (NSERC).

\section*{Ethics approval and consent to participate}

Not applicable.

\section*{Consent for publication}

Not applicable.

\section*{Availability of data and materials}

The source code and data are available at: \url{https://github.com/AEVO-lab/cnp2cnp}.

Supplementary file \textbf{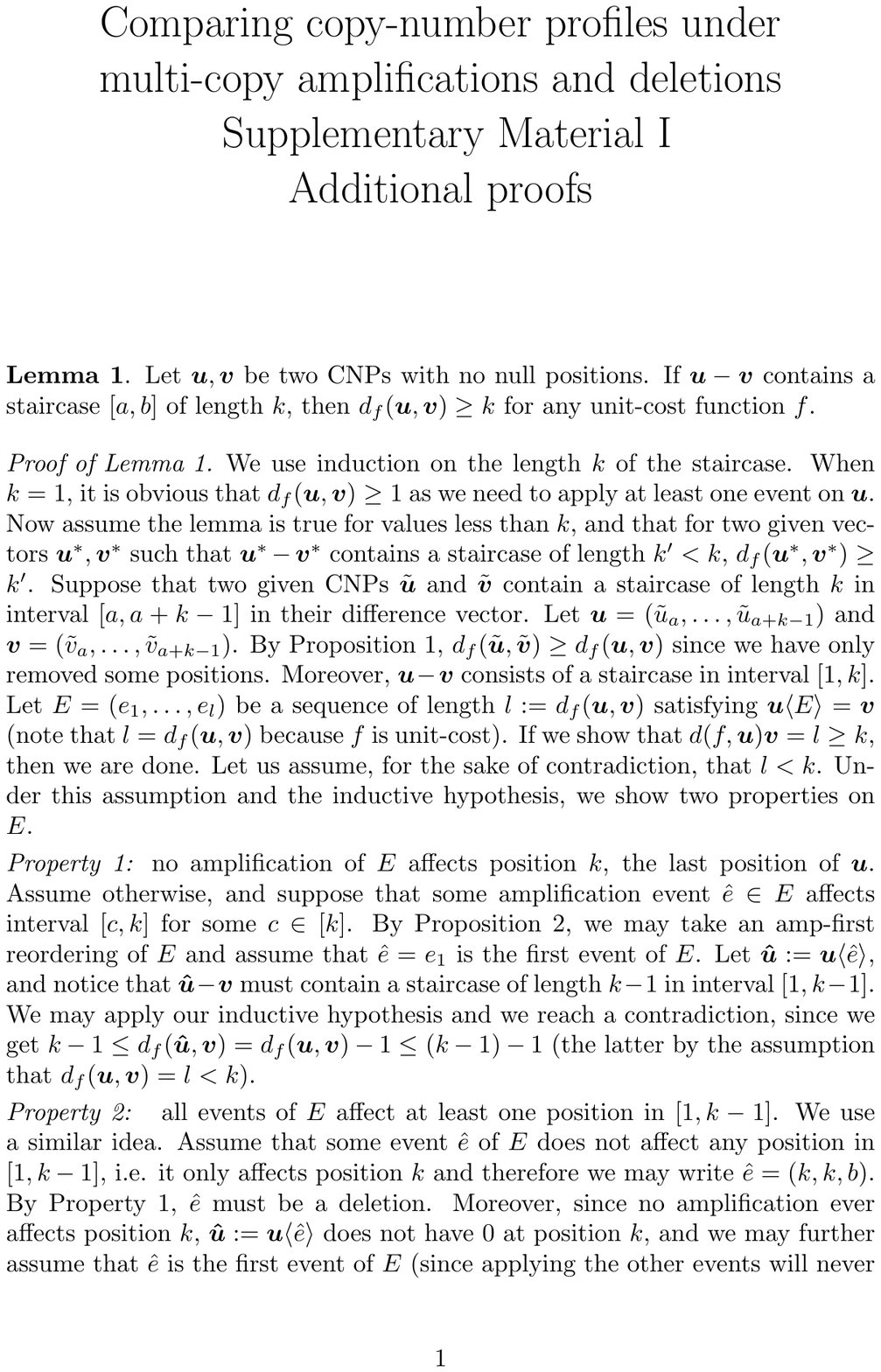} contains all the missing proofs.

Supplementary file \textbf{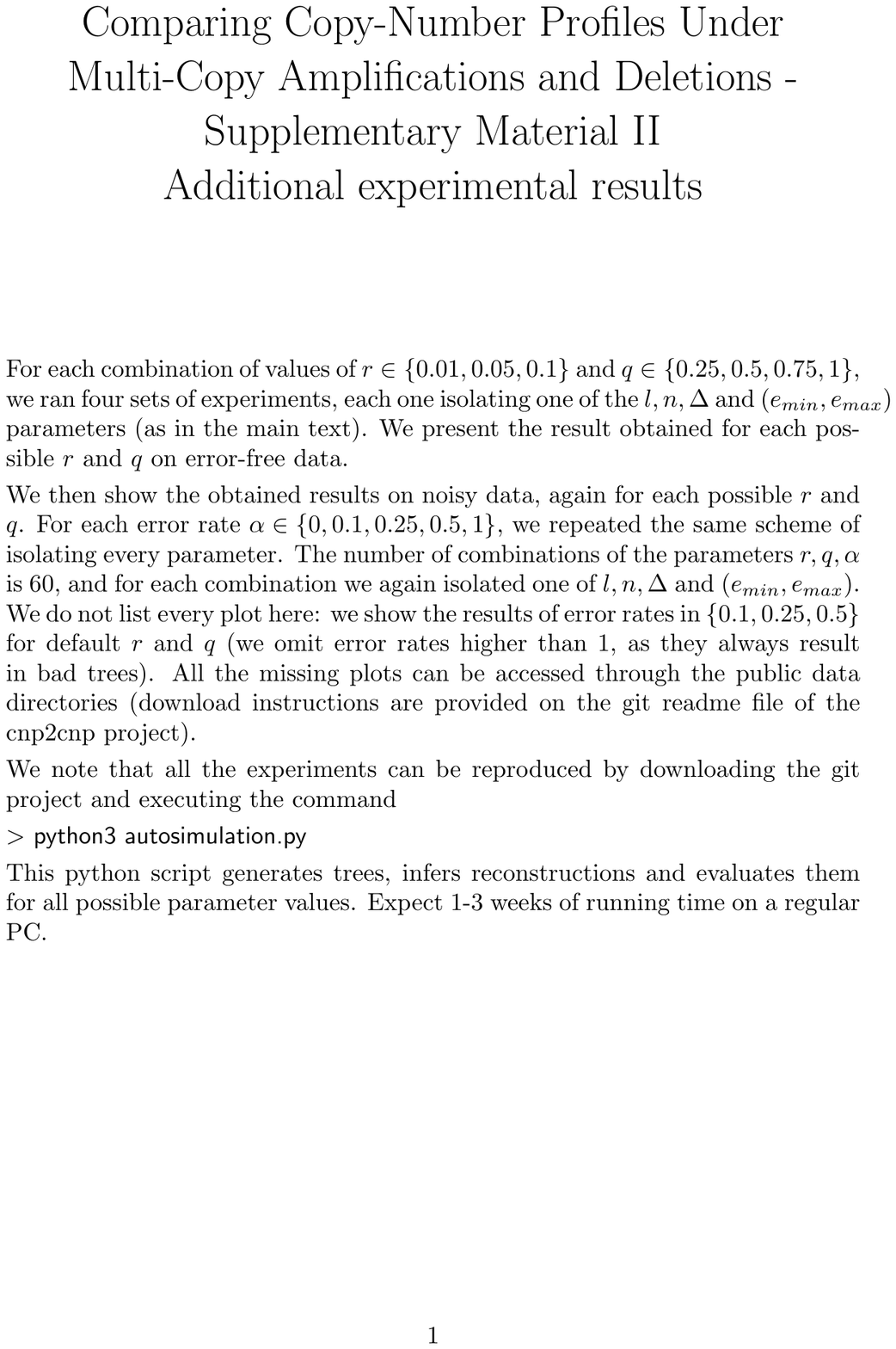} contains all the additional experimental results.


\end{backmatter}

\bibliographystyle{vancouver}

\bibliography{main}

\includepdf[page=-]{S33-S1}

\includepdf[page=-]{S33-S2}

\end{document}


\title{Comparing copy-number profiles under multi-copy amplifications and deletions \\\ Supplementary Material I \\
Additional proofs}

\date{}

\maketitle



\textbf{Lemma~\ref{lem:staircase}}.
Let $\u, \v$ be two CNPs with no null positions.
If $\u - \v$ contains a staircase $[a, b]$ of length $k$, then $\distf{f}{\u}{\v} \geq k$ for any unit-cost function $f$.

\begin{proof}[Proof of Lemma~\ref{lem:staircase}.]
We use induction on the length $k$ of the staircase.
When $k = 1$, it is obvious that $\distf{f}{\u}{\v} \geq 1$ as we need to apply at least one event on $\u$.  
Now assume the lemma is true for values less than $k$, and that for two given vectors $\u^*, \v^*$ such that $\u^* - \v^*$ contains a staircase of length $k' < k$, $\distf{f}{\u^*}{\v^*} \geq k'$.  Suppose that two given CNPs $\tilde{\u}$ and $\tilde{\v}$ contain a staircase of length $k$ in interval $[a,a+k-1]$ in their difference vector.  
Let $\u = (\tilde{u}_a, \ldots, \tilde{u}_{a+k-1})$ and $\v = (\tilde{v}_a, \ldots, \tilde{v}_{a+k-1})$.  By Proposition~\ref{prop:remove-stuff}, $\distf{f}{\tilde{\u}}{\tilde{\v}} \geq \distf{f}{\u}{\v}$ since we have only removed some positions.
Moreover, $\u - \v$ consists of a staircase in interval $[1,k]$.
Let $E = (e_1, \ldots, e_l)$ be a sequence of length $l := \distf{f}{\u}{\v}$ satisfying $\u\ev{E} = \v$ (note that $l = \distf{f}{\u}{\v}$ because $f$ is unit-cost).  If we show that $\dist{f}{\u}{\v} = l \geq k$, then we are done.
Let us assume, for the sake of contradiction, that $l < k$.  
Under this assumption and the inductive hypothesis, we show two properties on $E$.

\emph{Property 1:} no amplification of $E$ affects position $k$, the last position of $\u$.  Assume otherwise, and suppose that some amplification event $\hat{e} \in E$ affects interval $[c, k]$ for some $c \in [k]$.  By Proposition~\ref{prop:any-order}, we may take an amp-first reordering of $E$ and assume that $\hat{e} = e_1$ is the first event of $E$.
Let $\hu := \u\ev{\hat{e}}$, and notice that 
$\hu - \v$ must contain a staircase of length $k - 1$ in interval $[1,k-1]$.  We may apply our inductive hypothesis   and we reach a contradiction, since we get $k - 1 \leq \distf{f}{\hu}{\v} = \distf{f}{\u}{\v} - 1 \leq (k - 1) - 1$ (the latter by the assumption that $\distf{f}{\u}{\v} = l < k$).

\emph{Property 2: } all events of $E$ affect at least one position in $[1, k-1]$.  We use a similar idea.  Assume that some event $\hat{e}$ of $E$ does not affect any position in $[1, k-1]$, i.e. it only affects position $k$ and therefore we may write $\hat{e} = (k, k, b)$.  By Property 1, $\hat{e}$ must be a deletion.  Moreover, since no amplification ever affects position $k$, $\hu := \u\ev{\hat{e}}$ does not have $0$ at position $k$, and we may further assume that $\hat{e}$ is the first event of $E$ (since applying the other events will never make position $k$ drop below $0$).  In other words, $\distf{f}{\u}{\v} = \distf{f}{\hat{\u}}{\v} + 1$.  But then $\hu$ has a staircase in interval $[1, k-1]$ and by the same arguments as above, $k - 1 \leq \distf{f}{\hu}{\v} = \distf{f}{\u}{\v} - 1 \leq (k - 1) - 1$, again a contradiction.

So far, we know that only deletions affect position $k$ (Property 1), and all these deletions also affect position $k - 1$ (Property 2).
Because $u_{k-1} - v_{k-1} < u_k - v_k$ and $v_{k-1} > 0$, this implies that some amplification event $\hat{e}$ must affect position $k - 1$ (otherwise, applying only the deletion events affecting position $k$ on position $k - 1$ would make position $k - 1$ drop below $v_{k-1}$).
Let us assume, again using Proposition~\ref{prop:any-order}, that $\hat{e}$ is the first event of $E$, i.e. $e_1 = \hat{e}$.  
We use the same trick for a third time.  That is, let $\hu := \u\ev{\hat{e}}$ and notice that $\hu$ has a staircase in interval $[1, k-1]$.  Once again we obtain $k - 1 \leq \distf{f}{\hu}{\v} = \distf{f}{\u}{\v} - 1 \leq (k - 1) - 1$.
This contradiction forces us to conclude that $l < k$ is false, which proves the lemma.
\end{proof}

\vspace{5mm}

\textbf{Lemma~\ref{lem:smooth-stairs}}.
Let $\u$ and $\v$ be two CNPs with no null positions and let $f$ be any unit-cost function.  If $\u - \v$ contains a staircase in interval $[1, k]$ and $\distf{f}{\u}{\v} = k$, then there exists a smooth sequence transforming $\u$ into $\v$.

\begin{proof}[Proof of Lemma~\ref{lem:smooth-stairs}.]
We prove the lemma by induction over $k$.  As a base case, the statement is easy  to see when $k = 1$ since a single step can only removed by a deletion, which is smooth.  So assume $k > 1$ and that for any $\u', \v'$ such that $\distf{f}{\u'}{\v'} = k - 1$ and such that $\u' - \v'$ have a staircase of length $k - 1$ in $[1, k-1]$, there is an optimal smooth sequence transforming $\u'$ into $\v'$.

Let $E$ be any sequence of $k$ events such that $\u\ev{E} = \v$.  If $E$ is smooth, then we are done so assume otherwise.
The proof is divided in two parts.  
Assuming the inductive hypothesis, we first show that there is an optimal sequence $\hat{E}$ containing only deletions such that $\u\ev{\hat{E}} = \v$.
These deletions are not necessarily smooth.  We complete the induction in a second step, where we convert this deletion sequence into a smooth one.
For the remainder of the proof, we will denote $\w := \u - \v$.

\vspace{3mm}

\noindent
\textbf{Part 1: proof that $\u$ can be transformed into $\v$ using only deletions.}
Assume that $E = (e_1, \ldots, e_k)$ contains some amplification, otherwise we are done proving our first step.
We first claim that only deletions affect positions $k$ to $n$, inclusively.  
To see this, assume on the contrary that $e_i = (a, b, \delta)$ is an amplification where $b \geq k$.  
By Proposition~\ref{prop:any-order}, we may assume that $e_i = e_1$.  But $\u\ev{e_1}$
still has a staircase in interval $[1, k]$, and by Lemma~\ref{lem:staircase}, $\distf{f}{\u}{\v} \geq k$.  This is a contradiction since $e_1$ should reduce the distance to from $\u$ to $\v$.  Hence our claim holds.

We now claim that, on the other hand, some amplification in $E$ affects position $k - 1$.  This is clearly true if every deletion affecting position $k$ also affects position $k - 1$.  Indeed, we have $w_{k-1} < w_k$ and without an amplification on $k - 1$  it would be impossible that position $k - 1$ becomes equal to $v_{k-1} > 0$.  Thus if we suppose that no amplification affects position $k - 1$, there must be some deletion $e_i = (k, h, d)$ that affects position $k$ but not $k - 1$, where here $h \geq k$.  Let $\u' := \u\ev{e_i}$.  Since no amplification affects any position in $[k, h]$, $\u'$ has no position with value $0$.  
Furthermore, $\u' - \v$ contains a staircase of length $k - 1$ at $[1, k-1]$ and it is clear that $\distf{f}{\u'}{\v} = k - 1$.  By induction, there is a (smooth) deletion sequence $E'$ such that 
$\u'\ev{E'} = \v$.  In that case, the sequence formed by $e_i$ followed by $E'$ transforms $\u$ into $\v$ and has only deletions, which is what we want. Thus we may assume that our claim saying that some amplificatio affects $k - 1$ holds.

Moving on, 
let $e_i = (a, k - 1, \delta)$ be an amplification in $E$ that affects position $k-1$ (but not $k$).
Our previous claims show that $e_i$ exists.  By Proposition~\ref{prop:any-order}, we may assume that $e_1 = e_i$. Let $\u' := \u\ev{e_1}$ and $\w' := \u' - \v$. Then $\w'$ has a staircase of length $k - 1$ in interval $[1, k-1]$ and $\distf{f}{\u'}{\v} = k - 1$.  
Moreover, the differences in value between the steps have not changed, except at position $a$.  Formally, for each $i \in [k - 1] \setminus \{a\}$, $w'_i - w'_{i-1} = w_i - w_{i-1}$ and $w'_a - w'_{a-1} = w_a - w_{a-1} + \delta$.

By induction, $\u'\ev{E'} = \v$ for some smooth deletion sequence $E' = (e'_1, \ldots, e'_{k-1})$.  Here for each $i \in [k-1]$, $e'_i = (i, b_i, w'_{i-1} - w'_{i})$ for some $b_i \geq k - 1$.  Let $(i_1, b_{i_1}, d_{i_1}), \ldots, (i_l, b_{i_l}, d_{i_l})$ be the deletion events of $E'$ that affect position $k$, $i_1 < i_2 < \ldots < i_l$.  We distinguish two cases.

\vspace{3mm}

\noindent
\emph{Case 1:} $a \notin \{i_1, \ldots, i_l\}$.  
Then the event $(a, b_a, w'_{a-1} - w'_{a})$ of $E'$ does not affect position $k$, meaning that $b_a = k - 1$ (by smoothness). 
Consider the sequence $E''$ obtained from $E'$ by replacing the event $(a, k-1, w'_{a-1} - w'_{a})$ by the event $(a, k-1, w_{a-1} - w_{a})$.  Since $\u'\ev{E'} - \v$ has a $0$ everywhere and $w'_{a} - w'_{a-1} = w_{a} - w_{a-1} + \delta$, it follows that $\u'\ev{E''} - \v$
has value $0$ everywhere, except at positions from $a$ to $k - 1$ where it has value $\delta$.  But then, the only difference between $\u$ and $\u'$ is that positions $a$ to $k-1$ are increased by $\delta$.  Thus $\u\ev{E''} - \v$ has a value of $0$ everywhere (and $\u$ never drops below $0$, due to the smoothness of $E'$).  This means that $\u\ev{E''} = \v$, which is a contradiction since $E''$ has $k - 1$ events.

\vspace{3mm}

\noindent
\emph{Case 2:}
$a = i_h$ for some $h \in [l]$.  Then the deletion of $E'$ starting at $a$ is $(a, b_a, -(w'_a - w'_{a-1})) = (a, b_a, w_{a-1} - w_{a} - \delta)$ and affects position $k$, i.e. $b_a \geq k$.
Consider the sequence $E''$ obtained from $E'$ by replacing the event $(a, b_a, w_{a-1} - w_{a} - \delta)$ by $(a, b_a, w_{a-1} - w_{a})$.
Then $\u'\ev{E''} - \v$ has a $0$ everywhere, except at positions from $a$ to $b_a$ where it has value $\delta$.  Also, $\u\ev{E''} - \v$ has a $0$ everywhere, except at positions from $k$ to $b_a$ where it has value $\delta$.  We can apply the deletion $(k, b_a, -\delta)$ to $\u\ev{E''}$ to obtain $\v$.  Since $E''$ has $k - 1$ events, this yields a sequence of $k$ deletions transforming $\u$ into $\v$.

This concludes the first part.  That is, we have shown that if our inductive hypothesis holds, then some deletion sequence of length $k$ transforms $\u$ into $\v$.

\vspace{3mm}

\noindent
\textbf{Part 2: construction of a smooth sequence.}
Now let $\hat{E} = (\hat{e}_1, \ldots, \hat{e}_k)$ be a sequence of $k$ deletions transforming $\u$ into $\v$, which exists by Part 1.  
Let $(1, b, \delta)$ be any deletion affecting position $1$.  Since $\hat{E}$ contains only deletions, it is safe to assume that $\hat{e}_1 = (1, b, \delta)$.  Let $\u' := \u\ev{\hat{e}_1}$ and $\w' := \u' - \v$.  If $-\delta < w_1$, then 
$\w'$ contains a staircase of length $k$ and we reach a contradiction since this implies $\distf{f}{\u'}{\v} \geq k$.  If $-\delta > w_1$, then $w'_1 < 0$ and position $1$ can never have the same value as $v_1$ since $\hat{E}$ has only deletions.  We deduce that $-\delta = w_1$.  

It follows that $\u'$ has a staircase of length $k - 1$ in positions $[2, k]$.  
No event of $\hat{E}$ can affect position $1$ after $e_1$, so we can ignore this position in $\u'$ and $\w'$.  That is, suppose we remove position $1$ from $\u'$ and $\v$, yielding two vectors $\u''$ and $\v'$ of length $n - 1$.  Let $\w'' := \u'' - \v'$.  Then $\w''$ has a staircase of length $k - 1$ in interval $[1, k - 1]$.  This allows us to use induction, so that there is a smooth sequence $\hat{E}''$ of length $k - 1$ transforming $\u''$ into $\v'$.  This easily translates into a sequence $\hat{E}'$ transforming $\u'$ into $\v$: we just ``shift'' every event to the right to account for position $1$ in $\hat{E}'$.  To be specific, we replace any event $(s, t, \epsilon)$ from $\hat{E}''$ by the 
event $(s + 1, t + 1, \epsilon)$ in $\hat{E}'$.  Since $\hat{E}''$ is smooth, 
then we can write $\hat{E}' = ( (2, b_2, \epsilon_2), \ldots, (k, b_k, \epsilon_k))$
where, for each $i \in \{2, \ldots, k\}$, $b_i \geq k$ and $d_i = w'_i - w'_{i-1}$.

We have not shown smoothness yet, because $\hat{e}_1$ might not affect the whole $[1,k]$ interval as we wish.  
If indeed $\hat{e}_1$ affects position $k$, i.e. if $b \geq k$, then it is easy to see that applying $\hat{e}_1$ followed by $\hat{E}'$ is a smooth sequence transforming $\u$ into $\v$.  Thus we may assume that $b < k$.  Observe that $w'_i - w'_{i-1} = w_i - w_{i-1}$ for all $i \in \{2,\ldots,k\} \setminus \{b + 1\}$, because $w'_{b+1} - w'_{b} = w_{b+1} - w_{b} + w_1$ (recall that $-\delta = w_1$).  Let $(b + 1, b', w_{b} - w_{b+1} - w_1)$ be the deletion of $\hat{E}'$ that starts at position $b$, where $b' \geq k$ by smoothness.  
Suppose that we replace it with the deletion $(b + 1, b', w_{b} - w_{b+1})$ in $\hat{E}'$, yielding an alternate sequence $\tilde{E}$.  Then $\u' \ev{\tilde{E}} - \v$ has a $0$ everywhere, except at positions $b + 1$ to $b'$ where it has value $w_1$.  This means that if in $\hat{E}$, we replace $\hat{e}_1$ by $\tilde{e} = (1, b', -w_1)$ and follow it by $\tilde{E}$, we obtain a sequence transforming $\u$ into $\v$.  Now, let $\tilde{\u} := \u\ev{\tilde{e}}$.  
If we remove position $1$ from $\tilde{\u}$ (recalling that $\tilde{u}_1 = v_1$) and from $\v$, we obtain a CNP with a staircase at $[1, k - 1]$.  Applying induction, we get a smooth sequence $\tilde{E}''$ which we can modify into $\tilde{E}'$ to make it applicable to $\u$ (just as we did from $\hat{E}''$ to $\hat{E}'$).  It is then straightforward to see that $\tilde{e_1}$ followed by $\tilde{E}'$ is a smooth deletion sequence turning $\u$ into $\v$.
\end{proof}

\textbf{Theorem~\ref{thm:hardness}}.
The \cnp~problem is strongly NP-hard for any deletion-permissive unit-cost function, even if the CNPs have no null positions.

\begin{proof}[Proof of Theorem~\ref{thm:hardness}.]
From a \textsf{3-partition} instance $S = \{s_1, \ldots, s_n\}$, construct $\u$ and $\v$ as follows.  First define $K := 100n$
and, for all $i \in [n]$, put 
$p_i := \sum_{j = 1}^i s_j$, the idea being that $p_i$ and $p_{i-1}$ differ by an amount of $s_i$.  
Then put $\v$ as a vector containing only $1$s.  For $\u$, construct it by adding one position at a time from left to right: first insert the values $i + 1 + Kp_i$ for $i = 1 .. n$, and then the values $i(Kt + 3) + 1$ for $i = m .. 1$.  That is, let

\vspace{-7mm}
\begin{align*}
\v &= (1,1, \ldots, 1) \\
\u &= (2 + Kp_1, 3 + Kp_2, \ldots, n + 1 + Kp_n,m  (Kt + 3) + 1, \ldots, (Kt + 3) + 1)    
\end{align*}


This can be done in polynomial time in $n$ (in particular, each $p_i$ is polynomial). 
Observe that we have

\vspace{-7mm}
$$\w = (1 + Kp_1, \ldots, n + Kp_n, m (Kt + 3), \ldots, Kt + 3)$$

In particular, $\w$ has a staircase in interval $[1, n]$, followed by a decreasing staircase in interval $[n+1, n+m]$.  By Lemma~\ref{lem:staircase}, we know that $\distf{f}{\u}{\v} \geq n$.  We will show that $S$ is a YES-instance to \textsf{3-partition} if and only if $\distf{f}{\u}{\v} = n$.


\noindent
($\Rightarrow$): 
Suppose that there exists $m$ triplets $S_1,\ldots, S_m$ such that $\sum_{s' \in S_i} s' = t$ for all $i \in [m]$.  We may assume that each $s_i \in S$ is distinguishable, so that for each $s_i$ there is a unique $k$ such that $s_i \in S_k$. 
We construct a sequence $E = (e_1, \ldots, e_n)$ of $n$ deletions such that $\u\ev{E} = \v$.  For each $i \in [n]$, put $e_i = (i, n + k, w_{i-1} - w_i)$, where $k$ if the unique integer such that $s_i \in S_k$.
Note that the $e_i$ events are allowed because $f$ is deletion-permissive (this is actually the only place where we need this assumption).
One can check that $E$ is a smooth deletion sequence and it is clear that positions $1$ to $n$ become equal to $1$ after applying $E$ on $\u$.  Now consider the events that end at position $n + k$, $k \in [m]$.  For each $s_i \in S_k$, there is such an event that decreases all the positions $n + 1$ to $n + k$ by $w_i - w_{i-1} = Ks_i + 1$.  We get $\sum_{s_i \in S_k}(Ks_i + 1) = Kt + 3$.  Since this is true for every position from $n + 1$ to $n + m$, the total decrease for a position $k \in [m]$ will be $\sum_{j = k}^m Kt + 3 = (m +1 - k)Kt+3$, which is exactly $w_{n + k}$.  Hence $\u\ev{E} = \v$.

\noindent
($\Leftarrow$):
Assume that $\distf{f}{\u}{\v} = n$.  Let $E = (e_1, \ldots, e_n)$ be an optimal sequence of events transforming $\u$ into $\v$.  By Lemma~\ref{lem:smooth-stairs}, we may assume that $E$ is smooth. 
Thus each $e_i$ is a deletion of the form $(i, b_i, w_{i-1} - w_i) = (i, b_i, - (Ks_i+1))$, where 
$b_i \in [n, n+m]$. 
Let us define $S_k := \{s_i : b_i = n+k \}$.  We claim that $\sum_{s_i \in S_k} (Ks_i + 1) = Kt+3$.  For $k = m$, this must be true since $w_{n + m} = Kt + 3$.  For $k < m$, we have the difference $w_{n + k} - w_{n + k + 1} = Kt + 3$.  This means that the deletions that affect position $n + k$ but not $n + k + 1$ (i.e. those with $b_i = n + k$) must incur a total decrease of exactly $Kt + 3$, as claimed. 
We now argue that $|S_k| = 3$ for each $k \in [m]$.  Notice that $\sum_{s_i \in S_k}(Ks_i + 1) = K\sum_{s_i \in S_k}s_i + |S_k| = Kt + 3$.  If $\sum_{s_i \in S_k}s_i = t$, then $|S_k| = 3$.  Otherwise, by isolating the $|S_k|$ term above, it is not hard to deduce that $|S_k| \geq K$.  However, this is impossible since $|S_k| \leq n$ but $K > n$.  We have therefore shown that $|S_k| = 3$, which in turn implies that $\sum_{s_i \in S_k}s_i = t$.Therefore $S$ is a YES instance. 
\end{proof}

\textbf{Lemma~\ref{lem:fovr2}}.
Let $\u, \v$ be two distinct CNPs with no null positions, and let $\w := \u - \v$. 
Then for any unit-cost function $f$, 
$\distf{f}{\u}{\v} \geq \lceil (|F_{\w}| - 1)/2  \rceil$.

\begin{proof}[Proof of Lemma~\ref{lem:fovr2}.]
We prove the Lemma by induction on 
$\distf{f}{\u}{\v}$.  As a base case, when $\distf{f}{\u}{\v} = 1$, then $F_{\w}$ has 3 flat intervals: the extreme ones and the flat interval that gets affected in the single event transforming $\u$ into $\v$ (recall that we have artificial positions $w_0 = 0$ and $w_{n+1} = 0$, which guarantee that there are always two extreme intervals plus another one somewhere in $[i1,n]$).  The statement is clearly true in this case, as $\lceil |F_{\w}| - 1)/2 \rceil = 1$.

Now assume that the Lemma holds for any pair of CNPs $\u', \v'$ satisfying $\distf{f}{\u'}{\v'} < \distf{f}{\u}{\v}$.
Let $E = (e_1, \ldots, e_k)$ be an optimal sequence of events such that $\u\ev{E} = \v$.  Let $\hu := \u\ev{e_1}$ and $\hw := \hu - \v$.  Let $e_1 = (c, d, x)$, where $x$ could be negative in case of a deletion.
Let $F'_{\w} = \{[a,b] \in F_{\w} : [a,b] \cap [c,d] \neq \emptyset\}$ be the affected flat intervals.  
Assume that $F'_{\w}$ has $l \geq 0$ intervals, say $F'_{\w} = \{[a_1, b_1], \ldots, [a_l, b_l]\}$, and that they
are ordered so that $b_i + 1 = a_{i+1}$ for each $i \in [l - 1]$.  

First consider $[a_i, b_i]$ with $2 \leq i \leq l - 1$.  Note that $[a_i, b_i]$ cannot be an extreme flat interval in $\w$.  We claim that 
$[a_i, b_i]$ must still be a non-extreme flat interval in $\hu$.  
To see this, observe that $\hw_{a_i - 1} = \w_{a_i - 1} + x$ and 
$\hw_{a_i} = \w_{a_i} + x$.  Since $\w_{a_i - 1} \neq \w_{a_i}$ by maximality, we have $\hw_{a_i - 1} \neq \hw_{a_i}$.  By a similar argument, $\hw_{b_i + 1} \neq \hw_{b_i}$.  And because all values in $[a_i, b_i]$ have changed by the same amount $x$, $[a_i, b_i]$ is a (maximal)  flat interval (note that we need the assumption of no null positions to argue that all positions change by the same amount).
Moreover, $[a_i, b_i]$ cannot be extreme.  If instead $[a_i, b_i]$ was in the extreme interval containing $w_0$, then  we would have $\hw_h = 0$ for all $0 \leq h \leq b_i$.  In particular, this would imply $\hw_{a_i - 1} = \hw_{a_i}$, contrary to what we just argued.  The same occurs if we assume that $[a_i, b_i]$ is part of the extreme interval containing $w_{n+1}$.

Now consider any flat interval $[a, b] \in F_{\w} \setminus F'_{\w}$. 
It is easy to see that $[a, b]$ is still a flat interval in $\hw$, unless perhaps if $b + 1 = a_1$ or $a - 1 = b_l$.  In these cases, it is possible that $\hw_{b} = \hw_{a_1}$ and/or $\hw_{a} = \hw_{b_l}$.  These have the effect of ``merging'' two flat intervals, effectively eliminating
$[a_1, b_1]$ and/or $[a_l, b_l]$ (note that the argument also holds when $[a_1, b_1]$ or $[a_l, b_l]$ become part of an extreme interval).  Since every flat interval except these two stays in $\hw$, it follows that $|F_{\hw}| \geq |F_{\w}| - 2$.  
Then using induction, 

\vspace{-5mm}

\[
\distf{f}{\u}{\v} - 1 = \distf{f}{\hu}{\v} \geq \left\lceil (|F_{\w}| - 3)/2 \right\rceil = \left\lceil (|F_{\w}| - 1)/2 \right\rceil - 1
\]

and it follows that $\distf{f}{\u}{\v} \geq \left\lceil (|F_{\w}| - 1)/2 \right\rceil$.
\end{proof}

\textbf{Lemma~\ref{lem:conseq-zero}}.
Suppose that $v_i = v_{i+1} = 0$ for some position $i$.  Then removing position $i$ or $i + 1$, whichever is smaller in $\u$, from $\u$ and $\v$ preserves the distance between $\u$ and $\v$.  
Formally, for any unit-cost function $f$, if $u_i \geq u_{i+1}$, then $\distf{f}{\u}{\v} = \distf{f}{\u^{-\{i + 1\}}}{\v^{-\{i + 1\}}}$.  Similarly if $u_{i+1} \geq u_i$, then 
$\distf{f}{\u}{\v} = \distf{f}{\u^{-\{i\}}}{\v^{-\{i\}}}$.

\begin{proof}[Proof of Lemma~\ref{lem:conseq-zero}.]
Assume that $u_i \geq u_{i+1}$ (the other case is identical). 
We know that $\distf{f}{\u}{\v} \geq \distf{f}{\u^{-\{i + 1\}}}{\v^{-\{i + 1\}}}$, by Proposition~\ref{prop:remove-stuff}. 
We consider the converse bound.  Take any sequence $E = (e_1, \ldots, e_k)$ of events transforming $\u^{-\{i + 1\}}$ into $\v^{-\{i + 1\}}$.  
Modify $E$ to transform $\u$ into $\v$ as follows: each event affects the same positions as before (including those that have shifted after reinserting $i + 1$), but we ensure that every event affecting position $i$ also affects position $i + 1$.  
To be formal, define $E' = (e_1', \ldots, e_k')$ as follows.  
If $e_i$ increases interval $[a, b]$ by $\delta$ (which is possibly negative), then make $e'_i$ increase interval $[a', b']$ by $\delta$, where 

\vspace{-5mm}

\[
a' = \begin{cases}
a &\mbox{ if $a \leq i$} \\
a + 1 &\mbox{ if $a > i$}
\end{cases}
\quad \quad \quad \quad
b' = \begin{cases}
b &\mbox{ if $b < i$} \\
b + 1 &\mbox{ if $b \geq i$}
\end{cases}
\]

Aside from the new position $i$ in $\u$ and $\v$, every position reaches the same value as before.  Also because $u_i \geq u_{i+1}$, position $i + 1$ reaches $0$ after applying $E'$ on $\u$.
\end{proof}

\textbf{Lemma~\ref{lem:outstanding-nulls}}.
Suppose $v_i = 0$ for some position $i$ and that
$w_{i - 1} \geq w_i$ or $w_{i + 1} \geq w_i$.  Then $\distf{f}{\u}{\v} = \distf{f}{\u^{-\{i\}}}{\v^{-\{i\}}}$ for any unit-cost function  $f$.

\begin{proof}[Proof of Lemma~\ref{lem:outstanding-nulls}.]
The proof is essentially the same as in Lemma~\ref{lem:conseq-zero}.  If, without loss of generality, $w_{i-1} \geq w_i$, we can take an event sequence from $\u^{-\{i\}}$ to $\v^{-\{i\}}$ and adapt it so that every event affecting position $i - 1$ also affects position $i$.  This guarantees that position $i$ drops to $0$.  We omit the technical details.
\end{proof}

\section*{Finding good events in time $O(n \log n)$}

We say that an event $e$ is \emph{good} if applying it on $\u$ reduces $|F_{\w}|$ by $2$.
Here we present the detailed version of our improved heuristic.  The main algorithm that follows transforms $\u$ into $\v$ by making calls to the $findGoodEvent$ subroutine, which is defined afterwards.

\vspace{3mm}
\begin{algorithm}[H]
 \KwData{vectors $\u, \v$}
 \KwResult{Find a sequence that transforms $\u$ into $\v$}
 compute $\w := \u - \v$\;
 initialize empty sequence $S$\;
 \For{$u \neq v$}{
  \eIf{findGoodEvent($\u, \v, \w$) returns $(i,j,x)$}{
        add $(i,j,x)$ to $S$\;
        \For{$k = i, ..., j$}{
            $u_k$ = $\max{u_k + x, 0}$
        }
  }{
  find the first flat interval $[i,j]$ with $w_i \neq 0$\;
  increase $u_i, \ldots u_j$ by $-w_i$\;
  add $(i,j,-w_i)$ to $S$\;
  }
 }
 \Return{S}
 \caption{Main algorithm}
\end{algorithm}
\vspace{3mm}

The algorithm $findGoodEvent$ below can be implemented in time $O(n \log n)$.  
Our goal is to find a range of values $[i,j]$ that verifies $w_i - w_{i-1} = w_j - w_{j+1} := - \delta$.  We further need that $\delta > 0$, or that $\delta < 0$ and $\forall k \in [i,j], u_k \geq -\delta$ : we can then apply the event $(i, j, \delta)$.
To achieve this, the idea is simply to scan $\w$ from left to right.  Each time we detect a change of $w_k - w_{k+1}$, we check if we encountered the same amount of change before at some position $k'$ (this is $-\delta$ in the algorithm).  If so, we can return the $k, k'$ pair since it can be part of a good event.  Otherwise, we map $\delta = w_{k+1} - w_k$ to position $k + 1$ to store the fact that $k + 1$ is the latest position that could be matched with a change of $\delta$.  The last line of the for loop ensures that if we match two positions $k' < k$, all positions in-between are sufficiently high to allow a deletion of amount $\delta$.  

\vspace{3mm}

\begin{algorithm}[H]
 \KwData{vectors $\u, \v, \w$}
 \KwResult{Find an event that reduces $|F_{\w}|$ by $2$}
 initialization of an empty dictionary $R$\;
 \For{$k = 1, ..., n-2$}{
  $\delta := w_{k+1} - w_k$\;
  \textbf{if} $\delta == 0$ \textbf{then} continue \;  
  \eIf{$-\delta \in R$}{
   \Return{$(R[-\delta], k, \delta)$}\;
   }{
   Set $R[\delta] = k+1$\;
   delete all the key/value pairs $(x, y)$ in $R$ with $u_k \leq x$\;
  }
 }
 \Return{no possible event}
 \caption{findGoodEvent}
\end{algorithm}

We argue two components: that $findGoodEvent$ does find a good event, if there is one, and that it can be implemented to take time $O(n \log n)$.

\noindent
\textbf{Proof that Algorithm $findGoodEvent$ returns an event $(i,j,\delta)$ that reduces $|F_{w}|$ by $2$ when it exists.}
Consider an output $(i,j,\delta)$. Due to the construction, we had $-\delta \in R$, which can only be inserted with $-\delta = w_{i} - w_{i-1}$ and $\delta = w_{j+1} - w_j$, so $w_{i-1} - w_{i} = w_{j+1} - w_j$, in which case it is easy to see that $F_{\w}$ is reduced by $2$. Furthermore, if $\delta < 0$ and we had some $k \in [i,j]$ with $- u_k > \delta$, the $k$-th iteration would have deleted $\delta$ from $E$. This means that $(i,j,\delta)$ is indeed an event that reduces $|F_{\w}|$ and does not make any $u_k$ drop to $0$.

Reciprocally, if there is an event $(i,j,\delta)$ to be found we want to prove that the algorithm returns something (not necessarily the same event). If the algorithm exits before iteration $j$, it returns some event that we have already proven must be correct. Let us assume that we do not exit the loop before iteration $j$ : we have added $-\delta$ at rank $i$, and it is still in $R$ because for every $k \in [i,j]$ we did not have $-\delta > u_k$ by hypothesis. Since $-\delta$ is in $E$ and $w_{j+1} - w_j = x$, the algorithm returns $(i,j,\delta)$.

\noindent
\textbf{Complexity.}
The complexity of $findGoodEvent$ depends on the following operations: we need to be able to test the existence of a value in a dictionary, to add a key/value pair and, a bit less usual, to filter all values lower than a certain amount (the last line of $findGoodEvent$). 
%
We can use a \emph{treap} structure (see~\cite{seidel1996randomized}), which is a form of binary search tree that allows to split the values higher and lower to a certain number in $\log n$ time. This gives us a total complexity of $\mathcal{O}(n\log(n))$.

\bibliographystyle{plain}

\bibliography{main}